\pdfoutput=1

\documentclass[11pt]{article}

\usepackage{naacl2021}

\usepackage{times}
\usepackage{latexsym}

\usepackage{amsthm}

\theoremstyle{definition}
\newtheorem{definition}{Definition}[section]
\usepackage{arydshln}

\usepackage{pifont}
\usepackage{xcolor}
\usepackage{graphicx}
\usepackage{multirow}
\usepackage{url}        
\usepackage{hyperref}   
\hypersetup{
    breaklinks=true,    
}
\usepackage{listings}
\usepackage{xcolor}
\definecolor{offwhite}{RGB}{250,250,245} 
\lstdefinestyle{SQLStyle}{
    language=SQL,
    basicstyle=\ttfamily\small,
    keywordstyle=\color{blue}\bfseries,
    commentstyle=\color{gray},
    stringstyle=\color{red},
    morekeywords={WITH, AS, SELECT, FROM, WHERE, AND, COUNT, GROUP, BY}, 
    showspaces=false,
    showtabs=false,
    tabsize=2,
    breaklines=true,
    backgroundcolor=\color{offwhite}
}

\usepackage[T1]{fontenc}

\usepackage[utf8]{inputenc}

\usepackage{microtype}

\title{LLM-SQL-Solver: Can LLMs Determine SQL Equivalence?}

\author{ \bf
Fuheng Zhao,
Jiayue Chen\thanks{Work done at UC Santa Barbara.},
Lawrence Lim,\\
\bf 
Ishtiyaque Ahmad\footnotemark[1],
Divyakant Agrawal, 
Amr El Abbadi\\
UC Santa Barbara \\
\texttt{\{fuheng\_zhao,jiayuechen,lawrenceklim,ishtiyaque,divyagrawal,elabbadi\}@ucsb.edu}
}

\begin{document}
\maketitle
\begin{abstract}
Judging the equivalence between two SQL queries is a fundamental problem with many practical applications in data management and text-to-SQL generation. While the research community has reasoned about SQL equivalence for decades, it poses considerable difficulties and no complete solutions exist. Recently, Large Language Models (LLMs) have shown strong reasoning capability in conversation, question answering and mathematics challenges. In this paper, we study if LLMs can be used to determine the equivalence between SQL queries under two notions of SQL equivalence (\textit{semantic equivalence} and \textit{relaxed equivalence}). To assist LLMs in generating high quality responses, we present two  prompting techniques: \textbf{Miniature \& Mull} and \textbf{Explain \& Compare}. The former technique is used to evaluate the semantic equivalence in which it asks LLMs to execute a query on a simple database instance and then explore if a counterexample exists by modifying the database. The latter technique is used to evaluate the relaxed equivalence in which it asks LLMs to explain the queries and then compare if they contain significant logical differences. Our experiments demonstrate that using our techniques, LLMs are promising tools to help data engineers in writing semantically equivalent SQL queries, however challenges still persist, and have better alignment with human preference on text-to-SQL generations than execution accuracy.
\end{abstract}

\section{Introduction}
Structured Query Language (SQL) it is the standard programming language to query data systems and it is one of the most popular programming language~\cite{sqlpopular}. Since SQL is not a procedural language, one may write different SQL queries that are semantically equivalent. As a result, determining the equivalence of SQL queries is a fundamental task in many practical important applications ranging from query optimizers in database systems~\cite{pirahesh1992extensible, bai2023querybooster, zhou2021learned, samwel2018f1} to evaluating the translation from user text-to-SQL~\cite{zelle1996learning, zhong2020grounded, khatry2023words, cidrsql}. However, it has been shown by the data management research community that checking the equivalence of two SQL queries is in general undecidable~\cite{abiteboul1995foundations}. While there are theoretical works to check equivalence under different assumptions~\cite{chandra1977optimal, sagiv1980equivalences, cohen1999rewriting, chu2018axiomatic, wang2022wetune} and implemented prototypes such as Cosette~\cite{chu2017cosette}, they only support limited SQL operations and such projects are no longer maintained due to lack of experts. For instance, Cosette does not support commonly found operations like \textit{EXISTS}, \textit{IN}, queries with constant expressions, etc.; and Cosette is no longer maintained due to the lack of staff. Query optimizers and query processing engines in data systems still primarily rely on heuristic query rewriting rules~\cite{begoli2018apache}. Exploring and writing these rules correctly require strong domain expertise and need many engineering hours.

Moreover, due to the difficulty in evaluating the equivalence of two SQL queries, Text-to-SQL literature~\cite{zhong2017seq2sql, yu2018spider, wang2019rat, lee2021kaggledbqa, li2023can} primarily use \textit{exact match} (exact SQL string match) or \textit{execution match} (compare the execution outputs) as the evaluation metrics. However, exact string match may overlook many equivalent SQL queries (there are many ways different ways to write the same SQL query), and \textit{execution match} may misclassify different SQL queries as equivalent due to the lack of counterexamples in the underlining database instance. We note that there have been a few attempts that explore using semantic equivalence as the evaluation metrics~\cite{zhong2020semantic}. SQL queries in text-to-SQL benchmarks often include constructs such as \textit{EXISTS}, \textit{UNION}, \textit{UNION ALL}, and database schema doesn't no exclude NULL values, making off-the-shelf SQL provers (e.g., Cosette~\cite{chu2017cosette} and WeTune~\cite{wang2022wetune}) unsuitable for direct use. Consequently, researchers have turned to fuzzy testing, which assesses query semantic equivalence by generating numerous fuzzy database instances. However, this approach imposes substantial computational and storage demands. Specifically, evaluating a thousand pairs of SQL queries requires at least 75 minutes and consumes about 3 GB of storage space for creating these fuzz database instances~\cite{zhong2020semantic}. 
This substantial data requirement arises from a common issue in execution-based approaches: incorrectly classifying semantically inequivalent queries as equivalent due to the \textbf{not having a counter-example included}. For example, when the underlying database lacks tuples matching the predicate and join conditions, two semantically inequivalent queries can both produce an empty tuple as output. As a result, these queries will appear to match, regardless of their actual differences.

Recent advancements in large language models (LLMs) have shown great capability in following instructions and in generating decisions that align with human judgment~\cite{bubeck2023sparks, chiang2023can, yan2023codescope, zhang2023gpt}. For instance, It has been shown that GPT models are capable of performing first order logic reasoning~~\cite{saparov2022language}, detecting implicit contents~\cite{huang2023chatgpt, zhang2023schema}, and solving complex logical problems~\cite{feng2023language, yao2023tree}. In this paper, we investigate whether LLM can act as a judge to determine the equivalence of SQL queries. Hence, in this paper, we would like to mainly answer the following questions: 
\begin{itemize}
    \item Can LLMs help data engineers in identifying SQL semantic equivalence?
    \item Should LLMs be used to evaluate text-to-SQL generation?
\end{itemize}

The paper is organized as follows: Section~\ref{sec:background} motivates the need of LLMs in determining SQL equivalence and discusses the background information of two notions of SQL equivalence and LLMs basics. Section~\ref{sec:proposedMethod} introduces two prompting methods, Miniature and Mull and Explain and Compare, to improve LLMs' ability to assess SQL queries semantic and pragmatic equivalence. Section~\ref{sec:eval} presents experimental results on assessing SQL equivalence using GPT-3.5-Turbo and GPT-4o respectively, applying our two proposed prompting techniques on three datasets from the Spider benchmark. Section~\ref{sec:relatedWorks} discusses the implications of the proposed LLM-SQL-Solver  for SQL query validation and text-to-SQL evaluation, while also highlighting potential areas of improvement for our approach. Finally, Section~\ref{sec:conclusion} summarizes our contributions and concludes this work.

\section{Background}
\label{sec:background}

\subsection{Motivations}
We first introduce a query rewriting error that happened in the real-world system, known as the COUNT bug~\cite{ganski1987optimization}. Here are two SQL queries Q1 and Q2 as shown below:

\begin{lstlisting}[style=SQLStyle]
Q1:
SELECT Parts.pnum FROM Parts
WHERE Parts.qoh = (SELECT COUNT(Supply.shipdate) 
  FROM Supply
  WHERE Supply.pnum = Parts.pnum 
  AND Supply.shipdate < 10);

Q2:
WITH Temp AS
  SELECT Supply.pnum, COUNT(Supply.shipdate) AS ct
  FROM Supply
  WHERE Supply.shipdate < 10
  GROUP BY Supply.pnum
SELECT Parts.pnum FROM Parts,Temp
WHERE Parts.qoh = Temp.ct
AND Parts.pnum = Temp.pnum;
\end{lstlisting}
One may think these two queries are equivalent, since they both return results based on the condition that \textbf{qoh} from \textbf{Supply} table equals to the counts of records in the \textbf{Supply} table when \textbf{shipdate} is less than 10; however they are different in how they handle the subquery. Suppose that there is a record $R$ in the \textit{Parts} table such that \textit{R.qoh} is zero and \textit{R.pnum} does not exist in the \textit{Supply} table. 
Then, Q1 will include record $R$ since \textit{R.qoh} is zero and selecting the count for \textit{R.pnum} from the \textit{Supply} table will also return zero. On the other hand, Q2 will not include record $R$ because the \textit{TEMP} result will not contain \textit{R.pnum}.

Hence, real-world data systems need tools to avoid such pitfalls. While proving the semantic equivalency between two SQL queries has been proved to be undecidable~\cite{trahtenbrot1963impossibility, libkin2004elements, sagiv1980equivalences}, we believe LLMs may offer new opportunities to help data engineers in scrutinizing their SQL optimization logic and potentially identify flaws accordingly. For example, when we presented the COUNT BUG to ChatGPT, it was able to correctly identify that the two given SQL queries were not equivalent, effectively pointing out the underlying issue. Highlighting the potential of LLMs to complement traditional approaches by offering new insights.

Moreover, in text-to-SQL generation, many natural language questions are inherently ambiguous~\cite{katsogiannis2023survey, Floratou2024NL2SQL}, and as a result, they can be interpreted and answered in multiple ways~\cite{min2020ambigqa}. From the user's perspective, different SQL queries that convey the same logical answer to a question may all be considered equivalent. However, the widely used \textit{execution match} metric, which expects a single correct SQL query, does not capture these ambiguities. Consequently, we are interested in exploring whether LLMs can be utilized to assess the quality of text-to-SQL generation.

Moreover, in text-to-SQL generation, many natural language questions are inherently ambiguous~\cite{katsogiannis2023survey, Floratou2024NL2SQL}, allowing for multiple valid interpretations and corresponding answers~\cite{min2020ambigqa}. From a user's perspective, different SQL queries that yield the same logical answer to a question are often considered equivalent. The primary objective of text-to-SQL systems is to align with human preferences and deliver outputs that meet user expectations. However, the commonly used \textit{execution match} metric, which assumes a single correct SQL query, fails to account for these ambiguities. For example, consider a ground truth SQL query like "SELECT shipdate FROM Supply" and a predicted SQL query such as "SELECT shipdate, COUNT(shipdate) FROM Supply". Although these queries produce different output tables (making them execution unmatched), they convey the same underlying information. In such cases, a user might find the predicted query acceptable, even though it fails the strict execution match criterion. The limitations of execution match highlights the need for alternative approaches to evaluate the quality of text-to-SQL generation. On the other hand, large language models (LLMs) are often trained using Reinforcement Learning with Human Feedback (RLHF), which aligns their outputs more closely with human preferences and judgments~\cite{bai2022training}. We are particularly interested in exploring whether LLMs can provide a more nuanced assessment by identifying and validating diverse SQL queries that share the same intent, thereby achieving better alignment with human preferences.

\subsubsection{Two Notions of SQL Equivalence}
Motivated from the query optimizers in data systems and evaluation metrics for text-to-SQL, we define two notions of SQL equivalence.

\begin{definition}[Semantic Equivalence~\cite{codd1970relational}]~\label{formal-view}
Two SQL queries Q1 and Q2 are said to be {\em semantically equivalent} if and only if there does not exist a database instance such that Q1 and Q2 return different outputs.
\end{definition}

\begin{definition}[Pragmatic Equivalence~\cite{baker2018translation}]~\label{human-view}
Two SQL queries Q1 and Q2 are said to be {\em pragmatic equivalent} if they produce outputs that convey the same practical or meaningful information for the intended user, even if the outputs differ in format, structure, or additional details.
\end{definition}

Semantic equivalence is a well understood concept in database community. In contrast to semantic equivalence, pragmatic equivalence\cite{baker2018translation} is originated in natural language translation communities, focuses on maintaining the intended meaning or effect of an expression rather than its literal structure. For example, in natural language translation, a direct translation of the English idiom 'It rains cats and dogs' into another language would fail to convey the intended meaning, as the phrase is nonsensical. Instead, the pragmatically equivalent translation would be 'rain heavily,' which conveys the same idea of intense rainfall in a way that is meaningful to a non-English-speaking audience, without referencing cats or dogs. Similarly, in the context of SQL, pragmatic equivalence considers whether two queries achieve the same user intent or provide comparable practical information, even if their outputs or structures differ. This concept is particularly relevant in text-to-SQL systems, where multiple distinct yet valid queries can effectively fulfill the same user intent.

Based on the semantic definition of SQL equivalence (Definition~\ref{formal-view}), the Count bug is indeed a flaw for query engines and should be remedied immediately. On the other hand, we argue that based on Definition~\ref{human-view}, which reflects the logical perspective and aligned with human perception, the two SQL queries in the Count Bug example are pragmatic equivalent. The fact that even experienced experts failed to identify this subtle difference indicates humans often recognize equivalence based on the more relaxed pragmatic notion. This observation motivates us to explore whether LLMs can serve as a more effective metric for evaluating text-to-SQL tasks, aligning better with human judgment.


\begin{figure*}
    \centering
    \includegraphics[width=1\textwidth]{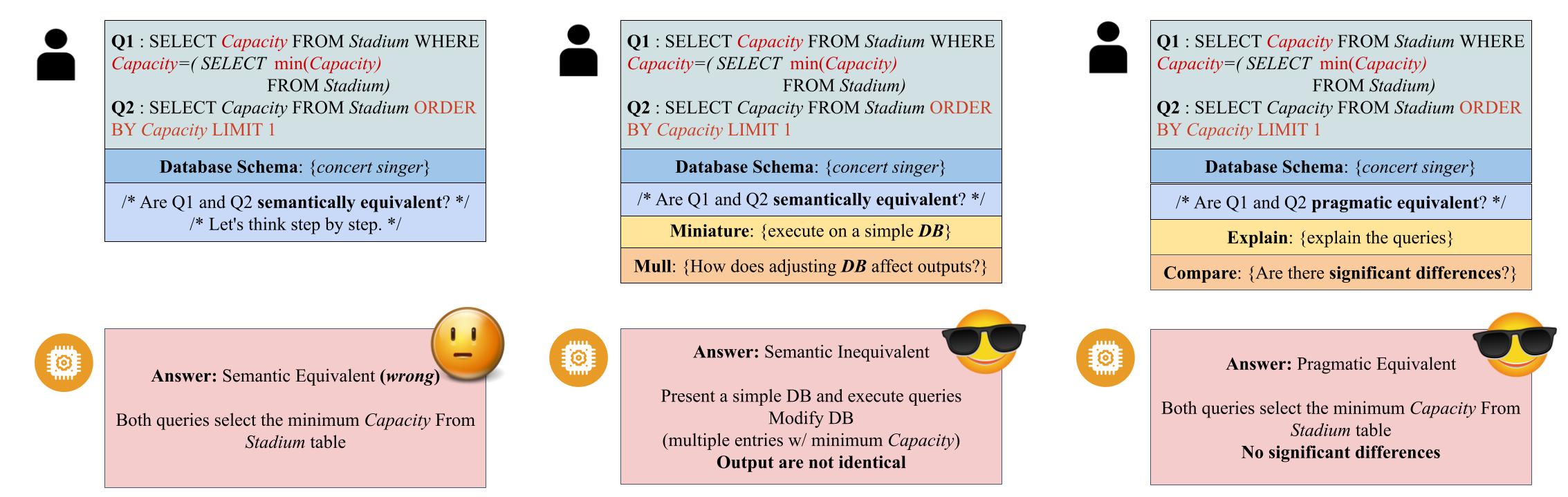}
    \caption{Demonstration of Chain-of-Thought, Miniature \& MULL, and Explain \& Compare prompts. The used SQL queries are not semantically equivalent but are pragmatic equivalent.}
    \label{fig:llm prompt}
\end{figure*}

\subsection{Large Language Models Basics}

Large language models (LLMs) are trained on extensive, web-scale datasets and leverage attention mechanisms~\cite{vaswani2017attention} combined with Reinforcement Learning with Human Feedback (RLHF) to generate high-quality, contextually relevant responses to input prompts. These models have demonstrated exceptional capabilities in solving complex tasks across diverse domains, such as natural language-to-SQL translation~\cite{pourreza2023din, gao2023text}, data system optimizations~\cite{zhou2023llm}, and data discovery~\cite{trummer2023can, fernandez2023large}.

A defining characteristic of LLMs is their ability to function as few-shot learners~\cite{brown2020language}. One widely used approach to enhance their accuracy is in-context learning (ICL). Through ICL, LLMs utilize example demonstrations provided within the input prompt to improve task performance dynamically. Morevoer, when prompted with specific instructions and requirements, such as generating output in JSON format, LLMs adapt their responses to align with these constraints, ensuring accurate and tailored results.


\section{Proposed Method}
\label{sec:proposedMethod}
Motivated by recent methods~\cite{deng2023rephrase, yao2023tree, wei2022chain}, we present our prompting techniques to obtain high quality judgments from Open-AI GPT models. Our prompting technique adheres to key aspects widely adopted in text-to-SQL works~\cite{li2023can, khatry2023words, liu2023comprehensive, taori2023stanford}: \textbf{Clear Layout}: We provide the two SQL queries that need to be compared at the beginning, then followed with the database schema, and end with the task (e.g., semantic equivalence or pragmatic equivalence) and instructions. \textbf{Clear Context}: We provide the database schema and explicitly included the primary and foreign key constraints, since they are crucial context information for deciding SQL equivalence. \textbf{Prompt style}: We adopted the code-representative approach, which has proven highly effective for database-related tasks~\cite{gao2023text}. An example prompt is illustrated below:

\begin{lstlisting}[style=SQLStyle]
/* Given the following SQL queries */
Q1:{Q1}
Q2:{Q2}
/* And the following database schema: */
{schema}
/* task and instructions */
\end{lstlisting}


For evaluating semantic equivalence, we propose two distinct prompts. One approach utilizes a chain-of-thought (CoT) prompt, where the task and instructions explicitly guide the model to determine semantic equivalence by reasoning step-by-step. While the chain-of-thought prompt is generally effective, as shown in Figure~\ref{fig:llm prompt}, it often yields suboptimal results. This is particularly evident when language models merely scratch the surface, focusing on obvious similarities in the structure and logic of SQL queries rather than nuanced differences. This observation aligns with prior studies~\cite{wang2023large, zheng2023judging}, which highlight that language models tend to overlook subtle distinctions when assessing equivalence and demonstrate limited reasoning capabilities in such contexts.

To overcome this observed phenomenon, we introduce our proposed prompting technique in detail to obtain better responses from large language models. In complex problem solving, people always start with a simple example and then introduce more complex elements to understand the full picture. Motivated by this simple intuition, we incorporate this idea and proposed the \textbf{Miniature and Mull} prompting technique. In particular, we first ask LLMs to execute both queries on a simple database. If the outputs of the two SQL queries are identical, we then instruct LLMs to adjust the simple database (e.g., updating rows) and observe how the adjustments affect the outputs.  If the modified database causes the two queries to produce different results, it indicates semantic inequivalence (i.e., a counterexample exists).

As shown later in the experimental section, this approach can directly enhance the LLM’s response accuracy in determining SQL semantic equivalence. We believe that this prompting technique has wide applications in comparing the equivalence of two codes, plans, and logic.

In addition, we have added a specific instruction in both \textit{CoT} and \textit{Miniature \& Mull} prompt: "The SQL predicate is case-sensitive when comparing string values". We have observed that LLMs often assume databases are case-insensitive. While many databases, such as SQL Server~\cite{mistry2014introducing} and MySQL~\cite{dubois2013mysql}, are indeed case-insensitive by default, the Spider benchmark employs SQLite, which is case-sensitive by default. To address this, both the \textit{CoT} and \textit{Miniature \& Mull} prompts explicitly specify that the database is case-sensitive when comparing string values.

For assessing pragmatic equivalence, we observed that LLMs often exhibit a tendency to label two SQL queries as equivalent based primarily on their structural and semantic similarity. This bias can lead to erroneous conclusions, as structural resemblance does not necessarily imply logical equivalence. To address this challenge, it is essential for LLMs to first identify and articulate the differences between the queries and then leverage this understanding to evaluate whether the queries are pragmatically equivalent. As a result, we designed the \textbf{Explain and Compare} prompt by asking the LLMs to perform multi-step reasoning before determining if the two SQL queries are equivalent. This prompt follows the main ideas from the chain-of-thoughts~\cite{wei2022chain}. We first prompt the LLMs to explain two SQL queries to understand their high-level objectives and then ask LLMs to determine if these two queries contain aligned logic or significant differences, as shown in Figure~\ref{fig:llm prompt}.
This structured approach mitigates the LLMs' inclination to focus solely on surface-level similarities. By requiring them to articulate reasoning steps, the prompt ensures a deeper evaluation of equivalence.

Moreover, we added specific instruction in \textit{Explain \& Compare} prompt regarding NULL values and case-sensitivity: "Database contains no NULL values". We have observed that LLMs often consider \textit{COUNT(*)}, \textit{COUNT(primary key)}, and \textit{COUNT(DISTINCT primary key)} to be different. The reason is that LLMs consider cases when NULL values can be inserted into a column with primary key constraint. In fact, the SQLite databases, by default, allow null values to be inserted to primary key column. However, in many database primary key constraints explicitly states that null values are not allowed~\cite{javadb}. Although LLMs correctly consider that having NULL values in the primary key column may lead to different results (note that \textit{COUNT(*)} considers null values as unique entries, where as \textit{COUNT(primary key)} ignores rows with null values), these are not significant issue and does not affect the meaning of the query.


\section{Evaluations}
We evaluate the capabilities of popular large language models in identifying semantic and relaxed equivalence between a pair of SQL queries. In particular, we showcase the experimental results using  the popular GPT 3.5 turbo (gpt-3.5-turbo) and GPT 4 turbo (gpt-4-1106-preview) models. The models are accessed via OpenAI API.

\textbf{Hyper-parameter:} We use zero demonstrations in our prompts. In addition, in the API call, we have set temperature to 0 which implies greedy sampling is used to generate the output; The max tokens are set to 3000; and the stopping tokens are set to "Answer:" for all experiments.

\subsection{Data Source}
\textcolor{black}{First, we compare our proposed Miniature\&Mull prompt with the CounterExample prompt on i) Non-equivalent Queries~\cite{chu2017cosette} and ii) Equivalent Queries~\cite{equivdata}}. \textcolor{black}{These two data sources include pairs of SQL queries that have been expertly labeled as semantically non-equivalent or equivalent.} 


\begin{itemize}
    \item \textbf{Nonequivalent Queries}~\cite{chu2017cosette}: Four challenging and one easy questions, including the COUNT BUG, one exam question from CSE344 at the University of Washington, two bugs users found on SQL solver, and a simple hello world example.
    \item \textbf{Equivalent Queries}~\cite{zhong2020semantic}: A hundred pairs of expert labeled equivalent queries. 
\end{itemize}


\textcolor{black}{Next, we showcase results on the alignment of our proposed LLM-SQL-Solver judgements with human preferences on semantic and relaxed notions using the Spider benchmark~\cite{yu2018spider}.}
The Spider benchmark contains 1034 development questions covering 20 databases in different domains. The Spider SQL queries are classified into four levels of complexity: easy, medium, hard, and extrahard (extra).
We evaluate the judgments of LLM-SQL-Solver against execution accuracy (EX) based on human preferences. In all experiments, we use the open sourced EX evaluation code from GitHub~\footnote{https://github.com/THU-BPM/ChatGPT-sql}. 

\begin{itemize}
    \item \textbf{Chat}~\cite{liu2023comprehensive}: Generate queries using ChatGPT with the zero-shot basic prompt.
    
    \item \textbf{DIN}~\cite{pourreza2023din}: Generate SQL queries by decomposing the task into fine grained sub-modules and use the GPT-4 model with few-shot prompting.
    
    \item \textbf{DAIL}~\cite{dail_sql}: The current SOTA open sourced technique on the Spider benchmark. It uses the GPT-4 model with few-shot prompting and incorporates self-consistency~\cite{wang2022self}.
\end{itemize}

\subsection{Judging Semantic Equivalence}

\begin{table}[tbph]
    \centering
    \begin{tabular}{l|l|l|l}
        {\bf Judge} & {\bf Prompt} & {\bf Noneq}  & {\bf Eq} \\ \hline
        GPT 35 turbo & CounterExample & .8 & .45 \\ 
        & Miniature\&Mull & 1.0 \textcolor{green}{$\uparrow$} & .57 \textcolor{green}{$\uparrow$} \\ \hline
        GPT 4 turbo & CounterExample & .6 & .7 \\ 
         & Miniature\&Mull  & 1.0 \textcolor{green}{$\uparrow$} &.83 \textcolor{green}{$\uparrow$} \\ \hline
        
    \end{tabular}
    \caption{Results in determining semantic non-equivalent and equivalent queries.}
    \label{ineql-res}
\end{table}

Recall, two queries can not be semantically equivalent if there exists a counter example (a database instance) such that two queries produce different outputs. We used two different prompting methods \textit{Counter Example} and \textit{Miniature \& Mull} to compare and contrast the accuracy of LLMs in determining semantic equivalence.

\begin{table*}[tbph]
    \centering
    \begin{tabular}{l||l|l| l l l l l}\hline
        {\bf Judge} & {\bf Data} & {\bf Equivalence/Prompt}& {\bf Easy} & {\bf Medium} & {\bf Hard} & {\bf Extra} & {\bf Acc} \\ \hline
        \hline
        - & Spider& -& 248 & 446 & 174& 166 & - \\ \hline
        
        EX & Chat & - & 221 & 331& 101& 72 & 70.1\% \\ 


         & DIN & - & 228  & 382 & 132 & 101  & 81.5\% \\ 
         
         & DAIL & - & 226  & 392 & 132 & 104  & 82.6\% \\ 
        \hline

        \hline
        GPT 3.5 turbo & Chat& Semantic/Miniature\&Mull  & 210 & 293& 84 & 48& 61.4\% \\ 

         & DIN & Semantic/Miniature\&Mull   &  207 & 307 & 96 & 76  & 66.3\% \\ 
         
         & DAIL & Semantic/Miniature\&Mull   &  224  & 317 & 85 & 95 & 70.4\% \\ 
        \hdashline

         & Chat & Relaxed/Explain\&Compare &  217  & 283 & 102 & 86 & 66.5\% \\

        & DIN &  Relaxed/Explain\&Compare & 223 & 360 & 133 & 114 & 80.3\% \\

         & DAIL & Relaxed/Explain\&Compare & 218  & 371 & 140 & 119 & 82.0\% \\ 
       
        \hline
         GPT 4 turbo & Chat& Semantic/ Miniature\&Mull  & 199 & 244 & 78 &  55& 55.7\% \\ 
         

          & DIN & Semantic/Miniature\&Mull  &  209 & 308  & 99 & 73 & 66.6\% \\

         & DAIL & Semantic/Miniature\&Mull  &  216 & 343  & 114 & 75 & 73.0\% \\ 
         
        \hdashline

         & Chat & Relaxed/Explain\&Compare &  221 & 358 & 116 & 98  & 76.6\% \\ 

         & DIN &  Relaxed/Explain\&Compare & 209 & 377 & 124 & 106 & 79.0\% \\ 

         & DAIL &  Relaxed/Explain\&Compare &   220 & 382 & 141 & 110 & 82.5\% \\ 
        \hline

    \end{tabular}
    \caption{A breakdown of the accuracy scores when using different metrics, i.e., EX and LLMs' judgements with varying prompt techniques. The compared SQL queries are the gold SQL queries from Spider and the generated SQL queries from Chat, DIN, and DAIL.}
    \label{table-nl2sql-eval}
\end{table*}


\textcolor{black}{As shown in Table~\ref{ineql-res}, we can observe that the \textit{Miniature \& Mull} prompt always provides a higher accuracy than the \textit{Counter Example} prompt.} When using the \textit{Counter Example} prompt (asking LLMs to determine SQL equivalence and provide a counter example if the SQL queries are not equivalent), GPT 4 turbo failed on a user found bug originally posted on Hackers News~\cite{hackernews}. 
This bug contains two SQL queries where one used \textit{DISTINCT} in selection and the other used \textit{UNION ALL} to union two sub-queries. Hence, GPT 4 turbo failed to recognize that \textit{UNION ALL} will not remove duplicate rows (though \textit{UNION} will). GPT 3.5 turbo failed on the CSE344 exam question which involves different sub-queries. When using \textit{Miniature \& Mull} (asking LLMs to first run a simple example over the given SQL queries and then examining if the example database can be modified to generate a counter example), GPT 3.5 turbo and GPT 4 turbo both successfully identified all the mistakes. \textcolor{black}{Our proposed \textit{Miniature \& Mull} prompt not only increased the accuracy of identifying non-equivalent SQL queries to 100\% for both models but also increased the accuracy for identifying equivalent SQL queries by at least 12\%.} As a result, we believe that letting LLMs execute SQL queries on a simple database instance and then assessing whether a counterexample can be generated by modifying the database is a more effective method to determine SQL semantic equivalence.



\subsection{Judging NL2SQL Generation}
There are two notions that one can use in judging the equivalence of two SQL queries, i.e., the semantic and relaxed notions. We use the proposed \textit{Miniature \& Mull} and \textit{Explain \& Compare} prompts to evaluate the equivalence of SQL queries following semantic and relaxed notions respectively.

In Table~\ref{table-nl2sql-eval}, we present the breakdown of the alignment between human preference and the \textit{execution match}/\textit{LLM judgement} on the SQL pairs (gold SQL from Spider and generated SQL from Chat, DIN, and DAIL). We observe that the relative ranking among Chat, DIN, and DAIL are preserved for both EX and LLM judgements. Since the generated SQL queries in Chat are from ChatGPT with zero-shot prompting, we expect the SQL queries from DIN and DAIL to achieve better accuracy. Since DAIL applied the self-consistency techniques (resource-intensive~\cite{dail_sql}), we expect DAIL to be more likely to produce the best answers. As shown in Table~\ref{table-nl2sql-eval}, the ranking among Chat, DIN, and DAIL based on execution accuracy, GPT 3.5 turbo judgments, and GPT 4 turbo judgements with varying prompts, all aligned with our expectation.

\subsubsection{Agreement between LLMs' Judgements and Execution Accuracy}
For a majority of the questions, LLMs' judgements and EX agree with each other. In particular, as shown in, Table~\ref{table-agreement}, agreements between LLMs and EX for the semantic equivalence notion have at least 69.6\%; similarly, the agreements between LLMs and EX for the relaxed equivalence notion have at least 82.4\%. We believe this is an encouraging signal that implies the good quality of the Spider benchmark and LLMs' judgements. The Spider benchmark contains high quality data points (1000 human hours spent~\cite{yu2018spider}). Creating these high quality databases, which is necessary for good EX, are expensive and does not scale. Assuming the underlying database is well designed and the EX is highly accurate, then a high agreement between EX and LLMs are obviously desirable.


\begin{table}[t]
    \centering
    \begin{tabular}{l|l|l|l}
        {\bf Judge} &  {\bf Equiv} & {\bf Data} & {\bf Agreement} \\ \hline
        GPT 3.5 turbo & Semantic & Chat & 69.6\%  \\ 
        &  & DIN &  72.6\%\\ 
         &  & DAIL & 74.6\%  \\ 
        \hdashline
        GPT 4 turbo & & Chat & 78.3\% \\ 
        &  & DIN & 78.9\% \\ 
         &  & DAIL &  84.6\% \\ 
        \hline
        GPT 3.5 turbo & Relaxed & Chat & 83.8\% \\ 
                &  & DIN & 88.4\% \\ 
         &  & DAIL & 90.0\% \\ 
        \hdashline
        GPT 4 turbo & & Chat & 83.8\% \\ 
                &  & DIN & 82.4\% \\ 
         &  & DAIL & 86.1\% \\ 

        
    \end{tabular}
    \caption{We compare the agreement between LLMs' judgements and execution matches.}
    \label{table-agreement}
\end{table}

\subsubsection{Human Evaluation on Samples}

Although, the agreements between execution accuracy and LLM's judgments are high, there are still lots of questions with disagreements. In this section, we provide a detailed analysis on the sampled questions from the set of questions where EX and LLM's judgement differs. Since there are hundreds of such questions (different results between EX and LLM) per data source and per GPT models, we sample 10 questions randomly from the SQL query pairs where LLMs and EX disagree.

\textbf{Equivalence Evaluated Manually}
To evaluate the equivalence between two SQL queries, we use the human preference approach. Three of the authors manually checked the equivalence of the sampled questions where LLMs and EX disagree. In addition, we 
have also differentiated the equivalence between the semantic notion and relaxed notion (in essence, the former is objective and the latter is subjective). To ensure the quality of human judgement, the annotations are conducted independently, and we use the majority vote to determine the label for each pair of SQL queries.

\begin{table}[t]
    \centering
    \begin{tabular}{l|l|l|l}
        {\bf Judge} &  {\bf Equiv} & {\bf Data} & {\bf Prefer} \\ \hline
        GPT 3.5 turbo & Semantic & Chat & 2/10   \textcolor{red}{$\downarrow$} \\ 
        &  & DIN & 4/10  \textcolor{red}{$\downarrow$} \\ 
         &  & DAIL & 5/10  - \\ 
        \hdashline
        
        GPT 4 turbo & & Chat & 10/10  \textcolor{green}{$\uparrow$} \\ 
        &  & DIN & 6/10  \textcolor{green}{$\uparrow$} \\ 
         &  & DAIL & 5/10  - \\ 
        \hline
        
        GPT 3.5 turbo & Relaxed & Chat & 6/10  \textcolor{green}{$\uparrow$} \\ 
        &  & DIN & 5/10  \textcolor{green}{$\uparrow$} \\ 
         &  & DAIL & 6/10  \textcolor{green}{$\uparrow$}  \\  
        \hdashline
        
        GPT 4 turbo & & Chat & 8/10  \textcolor{green}{$\uparrow$} \\  
                &  & DIN & 6/10  \textcolor{green}{$\uparrow$} \\ 
         &  & DAIL & 7/10  \textcolor{green}{$\uparrow$} \\ 

        
    \end{tabular}
    \caption{We compare the human preference between LLM judge and EX on 10 randomly sampled questions from different data sources.}
    \label{table: human alignment}
\end{table}

\textbf{Semantic Equivalence} 

When using the semantic equivalence notion, we find GPT 3.5 turbo may not deliver better judgment than EX, but GPT 4 turbo results in better alignments. When EX detects discrepancy in the SQL queries' outputs, it often indicates these two SQL queries are not semantically equivalent. There are a few rare cases that we observed in which the SQL queries are semantically equivalent but EX determines not equivalent. For example, when the provided database in Spider contains column values that can not be decoded (similar issues are also raised~\cite{spiderissue}, semantically equivalent queries can be misclassified by EX. 

\begin{figure}
    \centering
    \includegraphics[scale=0.8]{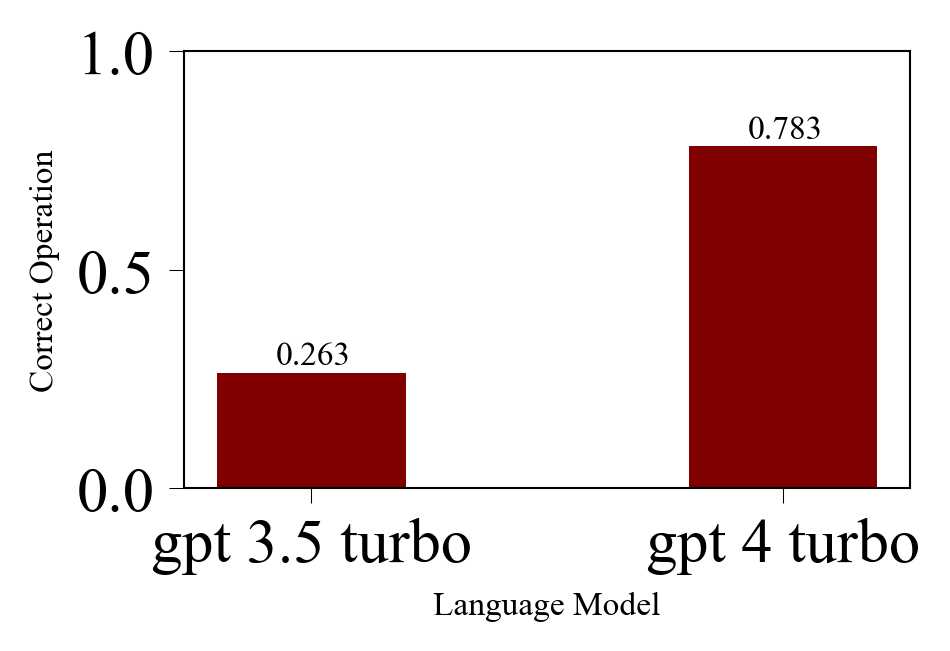}
    \caption{The rates of LLM correctly executed the SQL queries on its own provided simple database.}
    \label{fig:correct-ops}
\end{figure}

In Table~\ref{table: human alignment}, GPT 3.5 turbo aligned with human experts in the range of two to five out of ten. Diving deeper into the questions which GPT 3.5 turbo wrongly classified the SQL queries, we find that there are common reasons for the wrong classifications: \textbf{Distinct entries vs duplicated entries}. GPT 3.5 turbo often failed to consider the difference between selecting unique rows and selecting rows that may be duplicated; \textbf{Subtle difference in column names}. When the gold SQLs and generated SQLs contain only subtle differences in the columns used such as \textit{group by id} v.s. \textit{group by student\_id}, GPT 3.5 turbo may fail to notice these differences; \textbf{Hallucination}. GPT 3.5 turbo may return wrong results when executing the SQL queries on its own created simple database instance and sometimes when the outputs are indeed identical, it may still state that the outputs are different. This hallucination snowballing effect~\cite{zhang2023language} is one of the top reason why GPT 3.5 turbo performs worse compared to GPT 4 turbo.

When GPT 4 turbo is used as the judge, it constantly produced more accurate or comparable results to EX, with at least 5 out of 10 questions for DAIL and at most 10 out of 10 questions for Chat. One interesting observation is that GPT 4 turbo correctly identified the difference between \textit{order by asc/desc limit 1} v.s. \textit{where col = select min/max(col)}. The former will always return one row, as limit 1 is used. The latter, however, may return many rows when these rows share the same min or max value. EX often failed to capture this difference. Although GPT 4 turbo's judgment is much better, GPT 4 turbo may still miss some minor differences such as similar but different column names (such as song\_name v.s. name), overlook the fact that extra columns are selected, and also sometimes hallucinates. As shown in Figure~\ref{fig:correct-ops}, we explored how well do LLMs execute and reason about the SQL queries on their own generated simple database. GPT 4 turbo has much higher correct operation rates which leads to much better responses compared with the weaker model. One common phenomena that lead to false answers in EX is due to the \textbf{return of the empty tuple}. When the underlying database does not contain tuples that match the predicate, gold SQLs and generated SQLs will always match, irrespective of how different they are. Another interesting observation raised from a few comparisons is about the \textbf{case sensitivity}. OpenAI GPT models consider databases are by default case insensitive; Spider benchmark, on the other hand, uses the SQLite databases which are case sensitive by default. However, many popular database are by default case insensitive such as SQL Server~\cite{mistry2014introducing} and MySQL~\cite{dubois2013mysql}. 
Therefore, we operate under the assumption that the underlying database instance is case-insensitive when evaluating semantic equivalence.

\textbf{Relaxed Equivalence}. When using the \textit{Explain \& Compare} prompt for determining the relaxed equivalence, LLMs are consistently better aligned with human judgments than EX, as shown in Table~\ref{table: human alignment}. The relaxed notion aims to illustrate the degree to which humans perceive two queries as being similar in their meaning, and mimics the opinions of users. The positive results on the alignment between LLMs' judgments with human preferences indicate optimizing accuracy for LLMs' judgments is a better optimization target than EX.

The main reasons for why LLMs' judgements are better aligned with human preference than EX is that LLMs can zoom into the query logic and determine whether the two SQL queries share the same logic and contain the same truth. Consider the following example, \textit{order by col asc/desc limit 1} v.s. \textit{where col = select min/max(col)}. LLMs can correctly identify that these two queries are selecting the rows with minimal or maximal value. Since EX strictly checks the outputs, it may sometimes return equivalent and other times return not equivalent. One interesting observation we found is about \textbf{null values}. GPT often consider \textit{COUNT(*)}, \textit{COUNT(primary key)}, and \textit{COUNT(DISTINCT primary key)} to be different. The reason is that GPT models consider that null values can be inserted into a column with primary key constraint. In fact, the SQLite databases, by default, allow null values to be inserted to primary key column. However, in many database primary key constraints explicitly states that null values are not allowed~\cite{javadb}. Based on the relaxed equivalence definition, we consider records in the primary columns to be not null. Although LLMs correctly consider that having null values in the primary key column may lead to different results (note that \textit{COUNT(*)} considers null values as unique entries, where as \textit{COUNT(primary key)} ignores rows with null values), we consider these not as a significant issue.

\section{Discussion and Future Works}
\label{sec:relatedWorks}
In this section, we would like to highlight the broader implications and potential integration of our proposed LLM-SQL-Solver. Moreover, Concurrent with our research, the chain-of-table approach~\cite{wang2024chain} independently explores a similar concept, allowing LLMs to form a reasoning path through table modifications to improve question answering from tabular data, akin to our \textit{Miniature \& Mull} prompt. However, while our technique focuses on determining query semantic equivalence through table creation and modification, their goal is to deepen the understanding of the tabular data itself.

\subsection{Applications of LLM-SQL-Solver}
First, LLM-SQL-Solver (semantic notion) can be directly used to find system bugs (e.g., COUNT BUG) and help engineers validate their SQL rewriting code bases. Second, since query optimizer relies on heuristic rules to find equivalent SQL forms, it may not explore all rewriting possibilities and hence, the rewritten query may lead to sub-optimal performance. We believe that during the SQL rewrite phase of the query optimization engine, the LLM-SQL-Solver can be directly used as a tool to help checking semantic equivalence, ensuring bugs (e.g., COUNT bug~\cite{ganski1987optimization}) do not occur and also explore more SQL rewrite possibilities which may lead to better query performance.

For text-to-SQL tasks, we believe that when comparing the gold SQL query (the correct answer) with the generated SQL query, the current most popular metric, \textit{execution match}, does not reflects the human perspective and fails to capture a user's opinion on the generated SQL. For instance, a common reasoning for failures in \textit{execution match} is that the generated SQL selects fewer/extra columns than the gold SQL. In fact, a recent work has explicitly added \textit{"do not select extra columns"} in its prompt to improve the execution accuracy~\cite{dong2023c3}. However, from a user's perspective providing extra columns may not reduce the quality of the answer. Also, there are many other similar reasons that can lead to false execution matches but not necessarily indicate a lower quality of the generated SQL query (such as the column order in selection, case sensitivity, duplicate entries, and columns in different tables share the similar contents). 
Moreover, execution matching fails to correctly identify logically different queries when the underlying database does not contain any counter examples, even when hundreds of human hours were consumed to create the benchmark~\cite{yu2018spider}. Lastly, a fundamental limitation of execution accuracy is its need to query the underlying database. This is particularly pertinent as the text-to-SQL service provider usually does not have ownership of the user's database and the user’s database may also contain sensitive information. 
As a result, it would be desirable to have a SQL solver determining the relax notion of equivalence without constructing sophisticated databases (save costs) and without accessing any stored data (ensure data privacy). Based on our extensive experiment evaluations, we believe that LLM-SQL-Solver (with pragmatic notion) can be used as a better metric for text-to-SQL benchmarks.

\subsection{Future Plans}
There are areas of improvement to this work. First, we find LLMs may hallucinate in executing the queries over simple databases. By equipping LLMs with tools such as Python pandas library, LLMs can reply on these tool for observing the execution results of SQL queries~\cite{zhuang2023toolqa}. By reducing the hullicination, we believe that our proposed \textit{Miniature \& Mull} prompt will have an increased accuracy in determining SQL semantic equivalence. 

In addition, our evaluations of pragmatic equivalence were conducted on a set of 70 questions, requiring experts to manually label the equivalence of each pair of SQL queries. This process is time-consuming, as even experienced SQL experts need approximately 10 minutes to thoroughly analyze and determine the equivalence of a single query pair. Despite the relatively small dataset size, we observed consistent trends in the alignment between LLM judgments and execution matches. Notably, GPT-4o's responses consistently demonstrated better alignment with human preferences compared to other models, showcasing its ability to better capture nuanced distinctions in query intent. Expanding the dataset would allow for a more robust analysis and further validation of these trends, providing deeper insights into the strengths and weaknesses of current models.

Moreover, our analysis revealed that pragmatic equivalence is often highly subjective, depending on the context and the user's interpretation of query intent. This subjectivity underscores the need for designing customizable evaluation metrics tailored to individual users. Such personalized metrics could adapt to specific user needs and preferences, ultimately improving the quality and relevance of text-to-SQL systems for diverse applications. Future research in this direction could significantly enhance the usability and practicality of these systems, enabling them to cater to a broader range of use cases and user requirements.


\section{Conclusion}
\label{sec:conclusion}
In this paper, we studied and answered two main questions: i) Can current LLMs be used to help engineers in checking semantic equivalences between SQL queries and ii) should we use LLMs to judge the quality of generated SQL queries for text-to-SQL benchmarks. To the best of our knowledge, this is the first work studying the ability of LLMs in understanding SQL queries and determining SQL equivalences. Through extensive evaluations and manually examining the quality of the answers based on the alignment with human preferences, we find current LLMs (especially GPT-4o), using our proposed prompt techniques, can assists engineers in determining semantic equivalent queries, but may yield incorrect responses. On the other hand, when comparing the generated SQL query with the gold SQL query with pragmatic equivalence notion, LLMs' responses constantly achieve high alignments with human preferences. Moreover, using our proposed LLM-SQL-Solver as the evaluation metric for text-to-SQL translation ensures data privacy by eliminating the need to access stored user data, and it reduces costs by removing the requirement to construct sophisticated databases, making the evaluation process more efficient and scalable.

\section*{Acknowledgment}
We thank Tom Shangguan for his assistance in organizing and managing the data files used in this research.

\bibliography{custom}
\bibliographystyle{acl_natbib}




\end{document}